\documentclass[11pt,twoside]{article}
\usepackage{graphicx}
\usepackage{asp2006}
\usepackage{epsf}
\usepackage{epsfig}
\usepackage{lscape}

\begin{document}
\title{Practical suggestions on detecting exomoons in exoplanet transit light curves}
\author{
Gy.M. Szab\'o$^{1,2}$,A.E. Simon, L. L. Kiss$^{1,3}$, Zs. Reg\'aly$^{1}$}
\affil{
$^1$Konkoly Observatory of the HAS, Budapest, Hungary\\
$^2$Dept. of Experimental Physics, University of Szeged, Hungary\\
$^3$School of Physics, University of Sydney, Australia
}    

\begin{abstract}

The number of known transiting exoplanets is rapidly increasing, which
has recently inspired significant interest as to whether they can host
a detectable moon. Although there has been no such example where the
presence of a satellite was proven, several methods have already been
investigated for such a detection in the future. All these
methods utilize post-processing of the measured light curves, and the
presence of the moon is decided by the distribution of a timing
parameter. Here we propose a method for the detection of the moon {\it directly in the
raw transit light curves.} When the moon is in transit, it puts its own
fingerprint on the intensity variation. In realistic cases, this
distortion is too little to be detected in the individual light
curves, and must be amplified. Averaging the folded light curve of
several transits helps decrease the scatter, but it is not the best
approach because it also reduces the signal. The relative position of
the moon varies from transit to transit, the moon's wing will appear
in different positions on different sides of the planet's transit.
Here we show that a careful analysis of the scatter curve of the folded 
light curves enhances the chance of detecting the exomoons directly.
\end{abstract}

\begin{figure}
\centering\epsfig{file=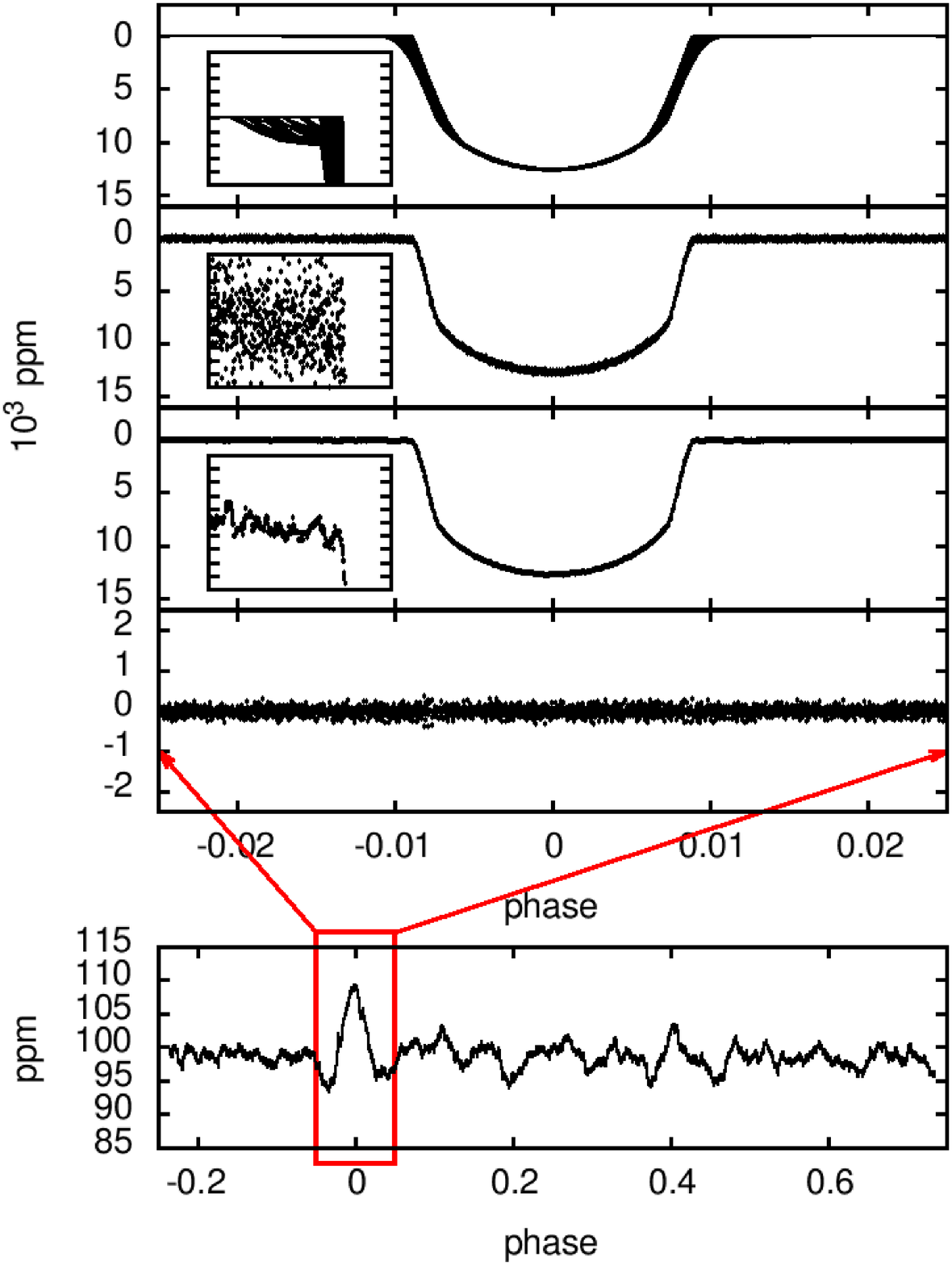,width=6cm}
\epsfig{file=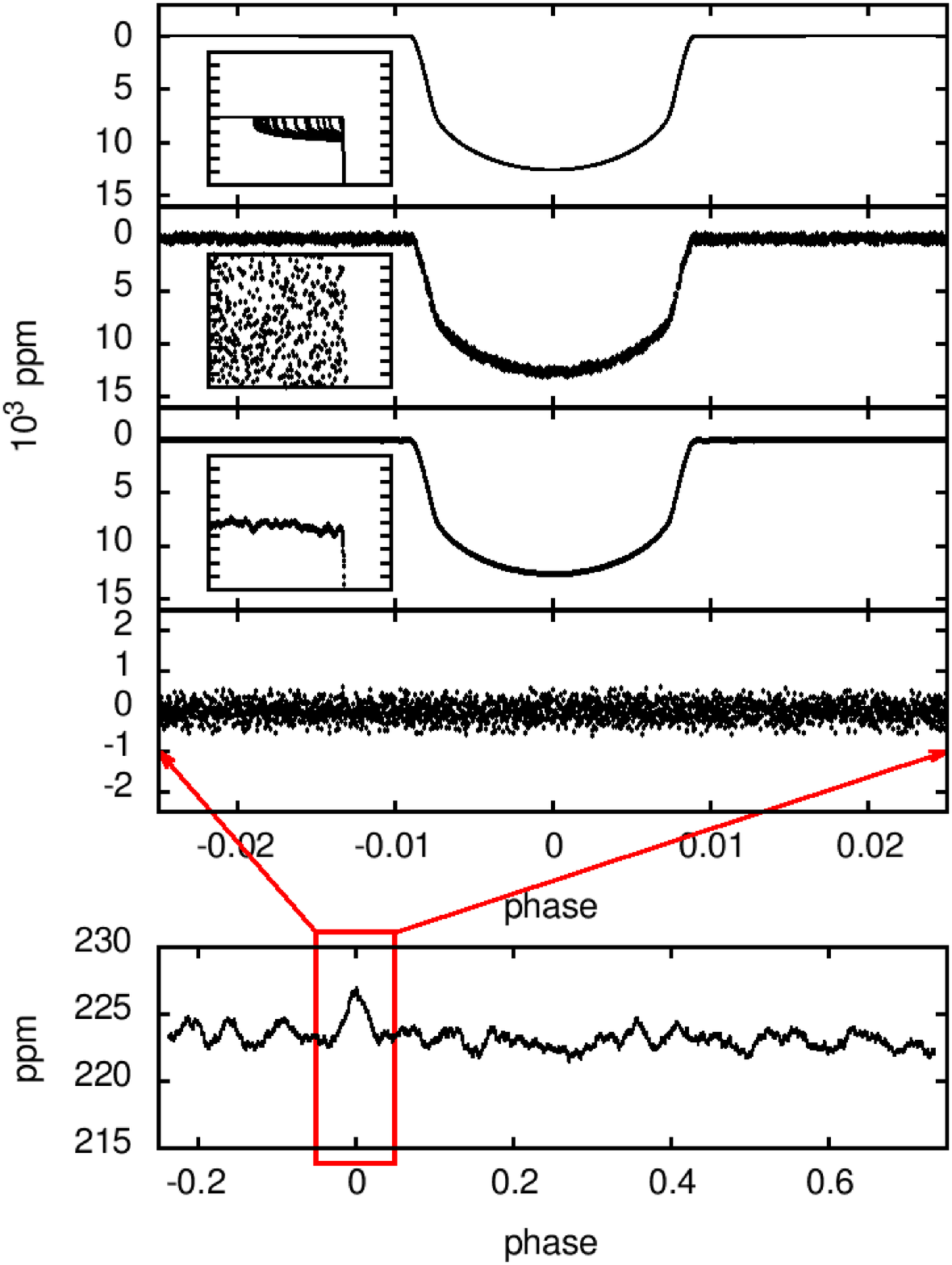,width=6cm}
\caption{Simulations of 110 transits with exomoon distortions, sampled as Kepler long cadence (left) and short cadence (right) data. Each column shows the input and the noisified light curves, the median filtered data, the residuals to the median and the rms scatter of the residuals. The inserts zoom to the exomoon signal (50 ppm per tick).}
\end{figure}

We simulated transit light curves of different qualities for exoplanets with moons. 
Space measurements covering 3 years were represented by
Kepler space telescope short cadence and long cadence samplings and the
bootstrap noise of non-variable stars.
Model moons had 0.7, 0.8, 0.9, 1.0 Earth-radius size, 
orbited on circular orbit with $P_{moon} = 4.3$~days, while the planet orbited
its circular orbit with $P_{planet}=10$~days period. The planet was a hot Jupiter with 
0.7 M$_J$, 1.0 $R_J$ mass and radius, while the mass of the moon was neglected.
The central star was a solar analog, the radius of orbits were $a_{planet}=0.09$~AU,
$a_{moon}=1.3\times10^6$~km, putting the moon to the border of the Hill-sphere.

We concluded that the most efficient tool of detecting the signal of the moon is observing
the residual scatter in the folded light curves, after subtracting a boxcar median of all
light curves. The scatter peak reflects the tiny light variations that occur because 
of the changing position of the moon in individual transits. We have deduced from simulations that for a clear detection (false alarm rate <1\%{}), 
cut levels in the 4.2--4.4$\sigma$ range must be chosen.
Selecting 4.4$\sigma$ treshold, 90\% of the Earth-sized moons
can be discovered in long cadence data, and practically all moons of this size will be
discovered in short cadence (detection rate is 99\%{}.) 
There is still some chance for the discovery of large exomoons with Earth-based quality 
observations via the scatter peak, in this case the discovery rate is 30\%{} for Earth-sized
moons.

\section{Practical suggestions for observations}

Testing the Scatter Peak from a sequence of light curves is
a promising tool for detecting moons directly in the light
curves. The successful detection relies on three important
conditions:

\begin{itemize}

\item{} All light curves must be stacked in such way that
the transit time of the planet exactly coincide
in every light curves involved to the analysis.
\item{} Transit observations must include the out-of
transit phases immediately before and after the transit of the
planet, where the scatter due to the moon is the
highest. The wings must be at least as long as
the transit duration.
\item{} Trend filtering the light curves must be done such
that the tiny brightness variations due to the exomoons
shall remain unaffected.

\end{itemize}

\acknowledgements
This work is supported by the Hungarian OTKA Grants K76816 and MB08C 81013, the ``Lend\"ulet'' Young Researchers' Program of the Hungarian Academy of Sciences and the Hungarian State ``E\"otvos'' Fellowship.

\end{document}